\documentclass[12pt,preprint]{aastex}

\slugcomment{Accepted for publication in ApJL}

\shorttitle{A double ring in Mrk~409}
\shortauthors{Gil de Paz et al$.$}

\begin{document}

\title{Discovery of a double ring in the dwarf galaxy Mrk~409}

\author{A. Gil de Paz\altaffilmark{1}, B. F. Madore\altaffilmark{1,2},
K. Noeske\altaffilmark{3}, L. M. Cair\'{o}s\altaffilmark{3}, P. Papaderos\altaffilmark{3}, S. A. Silich\altaffilmark{4}}

\altaffiltext{1} {The Observatories, Carnegie Institution of
Washington, 813 Santa Barbara Street, Pasadena, CA 91101; agpaz,
barry@ociw.edu}

\altaffiltext{2} {NASA/IPAC Extragalactic Database, California
Institute of Technology, MS 100-22, Pasadena, CA 91125}

\altaffiltext{3} {Universit\"ats-Sternwarte G\"ottingen, 37083,
G\"ottingen, Germany; knoeske, luzma, papade@uni-sw.gwdg.de}

\altaffiltext{4} {Instituto Nacional de Astrof\'{\i}sica, \'{O}ptica y
Electr\'{o}nica, AP 51, Luis Enrique Erro 1, Tonantzintla 72000,
Puebla, Mexico; silich@inaoep.mx}

\begin{abstract}
We report the discovery of a double ring of emission-line regions
around the nucleus of the Blue Compact Dwarf (BCD) galaxy Mrk~409 as
seen by deep, ground-based H$\alpha$ images. Echelle spectroscopy
obtained at Magellan-I with MIKE shows the presence of ionized gas
flowing out of the galaxy from a very massive
($\geq$7$\times$10$^{6}$\,M$_{\odot}$) nuclear starburst with
projected expansion velocities of $\sim$50\,km\,s$^{-1}$. Different
scenarios for the formation of these rings are discussed. While the
innermost, nuclear ring is most probably formed by the interaction of
a starburst-driven shock with the surrounding interstellar medium, the
origin of the outer ring is less clear.
\end{abstract}

\keywords{galaxies: evolution -- galaxies: starburst -- galaxies:
dwarf -- galaxies: individual(Mrk~409, NGC~3011) -- galaxies: kinematics and
dynamics}

\section{Introduction}
\label{intro}

Now in the era of the precision cosmology, where fundamental
parameters (H$_0$,$\Omega_{\mathrm{M}}$,$\Omega_{\Lambda}$) have been
measured with unprecedented accuracy (Bennett et al$.$ 2003), the
understanding of the evolution of the baryonic matter becomes even
more important than in the past. In this sense, the feedback of gas
and metals collectively ejected by starburst galaxies is currently the
most poorly-known ingredient of state-of-the-art hierarchical models
of galaxy formation and evolution (Kay et al$.$ 2002; Somerville \&
Primack 1999). This is particularly important in the case of dwarf
galaxies because of their shallow gravitational potentials. Despite
recent progress, the actual efficiency of this mechanism is still
unknown, mainly due to our poor understanding of the initial,
pre-burst galaxy state (Silich \& Tenorio-Tagle 2001; Mac Low \&
Ferrara 1999).

In this Letter we report the discovery of an unusual system of two
concentric rings of ionized gas around a massive nuclear starburst in
Mrk~409. Echelle spectroscopy obtained with MIKE at the 6.5-m
Magellan-I telescope at Las Campanas (Chile) shows ionized gas flowing
out of the disk of the galaxy. These results lead us to suggest that
the peculiar morphology of this system may be the result of the
interaction of a collective starburst-driven shock with the galaxy's
interstellar medium (ISM). Understanding the formation of these
peculiar structures may provide fundamental clues for the study of the
evolution of starbursts and their impact on the future evolution of
dwarf galaxies.

Mrk~409 is a nearby BCD galaxy
(v$_{\mathrm{helio}}$=1527\,km\,s$^{-1}$) with $B$-band luminosity
$M_B$=$-$17.27$\pm$0.03 ($h$=0.7; see Gil de Paz et al$.$ 2003; GMP
hereafter). It has been recently included in the Nearby Field Galaxy
Survey of Jansen et al$.$ (2000) and it was identified as a possible
counterrotator by Kannappan \& Fabricant (2001; KF01 hereafter).

\section{Observations and Reduction}
\label{observations}

On May 14-15, 2002 we obtained optical $BR$ and narrow-band H$\alpha$
($\Delta$$\lambda$$\simeq$20\AA) images of Mrk~409 at the Palomar
60-inch telescope using a 2048$\times$2048 CCD. Additional $BR$ images
were taken with the 1.23-m and 2.2-m telescopes at Calar Alto
(Almer\'{\i}a, Spain) on February 4, 2000 and March 10, 1997,
respectively. Deeper, higher quality $R$ and H$\alpha$
($\Delta$$\lambda$$\simeq$50\AA) images were later obtained (Jan 09
2003) using the ALFOSC camera at the 2.5-m NOT telescope (La Palma,
Spain) again using a 2048$\times$2048 CCD. Exposure times were 960\,s
($R$) and 7200\,s (H$\alpha$) and the seeing ranged between
0.9-1.0\arcsec. Finally, near-infrared ($JHK$) images were obtained at
the 1.5-m Carlos S\'anchez telescope (Tenerife, Spain) on March 8,
2000 using the CAIN camera with exposure times of 960\,s ($J$),
1140\,s ($H$), and 1440\,s ($K$). Images were reduced using standard
procedures of bias subtraction and flat-fielding within IRAF. Figure~1
shows the different images obtained and the corresponding surface
brightness and color profiles.

An echelle spectrum of Mrk~409 was obtained using MIKE at the
Magellan-I (Baade) telescope on Las Campanas (Chile) on April 28,
2003. For a complete description of MIKE see Bernstein et al$.$
(2003). The 1\arcsec-wide (5\arcsec-long) slit used provided a
spectral resolution of $\sim$14.1$\pm$0.4\,km\,s$^{-1}$ (FWHM) in the
light of H$\alpha$. The position angle of the slit was 39$^{\circ}$
and the exposure time 1200\,s (see Fig$.$~1b). The reduction and
extraction of the spectrum was carried out using the IRAF task {\sc
doecslit}. In Figure~2 we show the bidimensional image in the region
of H$\alpha$ (Panel~a) and the extracted spectrum in some of the lines
detected (b-f).

\section{Results and Discussion}
\label{results}

\subsection{Nuclear Starburst}
\label{starburst}

The center of Mrk~409 is of high surface brightness in the optical
($\mu_{B}$=19\,mag/$\sq\arcsec$; see Fig$.$~1e). This value is two
magnitudes brighter than the value obtained from the extrapolation of
the surface brightness profile of the underlying stellar population
(21.16\,mag/$\sq\arcsec$), which dominates the profile in the outer
regions of the galaxy. The enhancement in its surface brightness
profile along with the presence of ionized-gas emission associated
with the galaxy nucleus indicates the presence of a nuclear starburst.

If we assume an exponential profile for the low-surface-brightness
(LSB) underlying stellar population the decomposition of the surface
brightness profile yields a total luminosity for the starburst of
$M_{B}$=$-$16.5\,mag. However, the slightly convex shape of the
surface-brightness profile at distances $>$28\arcsec\ suggests a
central intensity flattening of the LSB component. Such flattening has
been recently found to be common within BCDs (see Noeske et al$.$
2003). This fact along with the moderately high extinction present in
the galaxy nuclear regions (inferred both from the galaxy nuclear
$B-R$ color and its Balmer decrement; see below; see also Jansen et
al$.$ 2000) make the number given above a lower limit to the
starburst luminosity.

The echelle spectrum of the galaxy nucleus reveals the presence of a
double-peaked H$\alpha$ emission-line with a total equivalent width of
11$\pm$1\,\AA\ (see Fig$.$~2b). These two peaks are shifted by
$-$54$\pm$3 and 37$\pm$3\,km\,s$^{-1}$ with respect to the radial
velocity of the stars. The latter was obtained by fitting the profile
of the MgI b 5184\,\AA\ absorption line (see Fig$.$~2c). 

Taking into account the contribution of the underlying stellar
population and the corresponding uncertainties the H$\alpha$
equivalent width of the nuclear starburst should be in the range
11-15\,\AA. If we now assume an instantaneous burst with a Salpeter
IMF (1-100\,M$_{\odot}$) and a typical metallicity of Z$_{\odot}$/5
the age derived would be $\sim$10.7-11.3\,Myr (Leitherer et al$.$
1999). However, as we show below, a significant fraction of this
emission is certainly arising from shock-excited gas. Therefore, since
the timescale for the injection of mechanical energy is much longer
than the photoionization timescale ($\sim$10\,Myr), the nuclear
starburst could be significantly older than this value. Thus, using
10\,Myr as a lower limit for the age we derive a total stellar mass
for the starburst $\geq$7$\times$10$^{6}$\,M$_{\odot}$.

The H$\alpha$ blue and redshifted components show very different
intensities. The approaching side is almost a factor of 2 brighter
than the receding one. This cannot be attributed just to a
dust-opacity effect since the H$\alpha$/H$\beta$ ratio of the
redshifted (4.86$\pm$0.25) and blueshifted (4.91$\pm$0.25) components
are comparable. The two components also show different excitation
conditions. The [NII]6583\,\AA/H$\alpha$ line ratio of the redshifted
component is [NII]/H$\alpha$$_{\mathrm{red}}$=0.480$\pm$0.007 while
for the blueshifted one
[NII]/H$\alpha$$_{\mathrm{blue}}$=0.306$\pm$0.005 (see
Fig$.$~2d,f). The [SII]6717,6731\,\AA/H$\alpha$ and [OIII]/H$\beta$
line ratios are log\,[SII]/H$\alpha$$_{\mathrm{blue}}$=$-$0.34 and
log\,[OIII]/H$\beta$$_{\mathrm{blue}}$=0.51 while\newline
log\,[SII]/H$\alpha$$_{\mathrm{red}}$=$-$0.12 and
log\,[OIII]/H$\beta$$_{\mathrm{red}}$=0.67. These numbers indicate
that the emission of both components, but particularly that of the
redshifted one, has a significant contribution from shock-excited gas
(see Martin 1997). One possible explanation is that the larger
contribution from photoionized gas to the blueshifted component is due
to its greater proximity to the ionization source. This would also
explain why this component is brighter than the one arising from the
more distant, shock-excited, redshifted gas. However, if that would be
the case we should expect the blueshifted lobe to be moving with a
lower velocity (with respect to the systemic velocity) than the
redshifted one, in contradiction with what we observe. 


It is also possible that either the amount of ionizing photons
absorbed by dust before ionization or the escape fraction of these
photons are significantly smaller for the blueshifted lobe than for
the redshifted one. This would result both in a larger contribution of
the photoionized-gas emission compared with that from
shock-excited-gas and a larger total emissivity of the blueshifted
lobe.

\subsection{Nuclear ring}
\label{inner}

At a distance of 5\arcsec\ (0.6\,kpc) from the galaxy nucleus we find
an H$\alpha$-emitting ring delineated by at least 8 individual H{\sc
ii} regions (see Fig$.$~1b). The mean H$\alpha$ equivalent width of
the regions in the ring as measured from the azimuthally-averaged $R$
and H$\alpha$ surface brightness profiles is 45$\pm$4\,\AA. The
equivalent width of H$\alpha$ corrected for the contribution of the
underlying stellar population is estimated to be in the range
70-80\,\AA\ (see Fig$.$~1f). Unfortunately, the continuum emission at
the position of this nuclear ring is also strongly contaminated by the
nuclear starburst, so EW(H$\alpha$) values higher than this number are
expected. With regard to the $B-R$ color, the position of the ring
coincides with the bluest portion of the galaxy profile. Higher
resolution images would be desirable in order to derive the properties
(color, luminosity, age, mass) of each individual region in the
ring. 

Interestingly, the distribution and properties of the regions in the
nuclear ring resemble those in the BCD Mrk~86. Like in Mrk~409, Mrk~86
shows a ring of young H{\sc ii} regions at a galactocentric radius of
0.5-1\,kpc around a massive, relatively evolved nuclear starburst. In
Gil de Paz et al$.$ (2000a,b) we proposed a scenario for the evolution
of Mrk~86 where the gas, swept out by the nuclear starburst in the
galactic plane, eventually reached densities large enough to become
molecular, thereafter forming a ring of H{\sc ii} regions and
newly-born stars. The similarity in the apparent properties of these
two galaxies suggests that the same mechanism could have taken place
in Mrk~409.

Using our numerical model for the evolution of a starburst-driven
bubble (Silich \& Tenorio-Tagle 1998) along with the evolution of the
kinetic-energy deposition rate predicted by the Leitherer et al$.$
(1999) models for a 7$\times$10$^{6}$\,M$_{\odot}$ massive starburst
we derive an average density for the surrounding ISM of
$n_{\mathrm{ISM}}$$\simeq$2\,cm$^{-3}$ and an evolutionary time of
$\sim$7.4\,Myr. Here we have adopted a shell radius of 600\,pc and an
expansion velocity of 50\,km\,s$^{-1}$. The small difference between
the bubble evolutionary time and the starburst age can be explained as
the time required for the coherent shell to form (see Silich et al$.$
2002).

The total H$\alpha$ luminosity of the shell (considering both photo-
and shock-ionized emission) can be written as
1.36$\times$(4$\pi$$R_{\mathrm{shell}}$$n_{\mathrm{ISM}}$$v_{\mathrm{shell}}$$+$$N_{\mathrm{UV}}$)\,10$^{-12}$\,erg\,s$^{-1}$,
where $N_{\mathrm{UV}}$ is the number of ionizing photons per second
escaping from the star-cluster region. Adopting an evolutionary time
for the bubble of 7.4\,Myr and the number of Lyman photons emitted by
the starburst as the value for $N_{\mathrm{UV}}$ the expected
H$\alpha$ luminosity would be
2.5$\times$10$^{40}$\,erg\,s$^{-1}$. This number is larger than the
total H$\alpha$ luminosity of the galaxy
(1.6$\times$10$^{40}$\,erg\,s$^{-1}$; GMP) suggesting that a
significant fraction of the ionizing photons are trapped within the
nuclear starburst region as it is also shown by Fig$.$~1e, and/or the
actual starburst age exceeds the adopted value.

It is worth mentioning that according to the simulations carried out
in Gil de Paz et al$.$ (2002) the ring of swept-out gas is not
necessarily in equilibrium and, in fact, has a rotational velocity
smaller than that of the underlying stellar population. This behavior
could explain the smaller velocity gradient of the gas compared with
the stars observed by KF01 in the inner regions of Mrk~409.

\subsection{Outer ring}
\label{outer}

Our deep H$\alpha$ image of Mrk~409 also reveals the presence of an
outer ring of line-emitting clumps at larger galactocentric distance
(see Fig$.$~1b). A small enhancement in the continuum light is also
observed in the broad-band images at a similar position (see
Fig$.$~1d,e). The overall distribution of the brightest clumps both in
the H$\alpha$ image and the unsharp-masked $B$-band image seems to be
well reproduced by an ellipse with ellipticity $\sim$0.1-0.2, position
angle $\sim$50$^{\circ}$, and major axis radius 19\arcsec\
($\sim$2\,kpc). Note that although the rings in Mrk~409 are named
according to the terminology used for lenticular and spiral galaxies
their formation mechanism may be different. In particular, Mrk~409
does not show any obvious oval distortion or bar, which are believed
to be responsible for the formation of rings in those galaxies (see
Buta \& Combes 1996).

A first explanation for the geometry observed in H$\alpha$ is that the
outer ring represents the brightened-limb of an hour glass-like
structure of ionized gas with the innermost ring being the narrowest
part of it, where the gas present in the galactic plane would have
been swept out (see Silich \& Tenorio-Tagle 1998). However, both the
enhancement in the continuum light associated with this ring (which
suggests the presence of secondary star formation) and the high
intensity contrast between the ring and the inter-ring regions allow
us to rule out this scenario. 

Therefore, in the case of the outer ring, an independent mechanism
must have been responsible for shaping this ring and triggering its
recent star formation activity. Some scenarios are proposed, (1) the
existence of an episodic starburst activity in the nucleus that has
resulted in the formation, first, of the outer ring and, more
recently, of the nuclear one, (2) the merging with a companion galaxy
or cloud, or (3) gas accretion from the galaxy halo.

The standard-model equations for the evolution of the bubble radius
and expansion velocity (Mac Low \& Mac Cray 1988) imply that, in order
for the scenario (1) to be compatible with the starburst mass and
outer-ring radius measured, the host galaxy has to have an extended
gaseous halo with an average density well below 1\,cm$^{-3}$.

On the other hand, the remarkably regular morphology of the galaxy LSB
component suggests that if the scenario (2) is correct the collision
probably involved a very low mass gas-rich companion galaxy, as those
proposed by Bergvall \& \"Ostlin (2002) to explain the recent star
formation activity in BCD galaxies.

Finally, the scenario (3) requires that the outer-ring radius roughly
coincides with the boundary of the radial velocity curve that shows
solid-body rotation. However, stellar spectroscopy carried out by KF01
indicates that the radius where the rotation curve flattens
($\leq$0.7\,kpc) is significantly smaller than the outer-ring radius
($\sim$2\,kpc).


\section{Conclusions}

We have discovered a double ring of H$\alpha$-emitting regions in the
BCD Mrk~409. Although multiple rings are common in lenticular and
spiral galaxies (Buta \& Combes 1996) this is the first time that such
a complex morphology is observed in a dwarf galaxy like
Mrk~409. Echelle spectroscopy reveals the presence of gas flowing out
of the galaxy most probably associated with a starburst-driven
supershell arising from a massive (relatively obscured) nuclear
starburst. The formation of the nuclear ring is interpreted as being
due to the sweeping of the ISM by the shock in the galactic plane.

The origin of the outer ring is less clear. Different scenarios are
proposed: (1) the existence of previous nuclear starburst activity
that shaped an expanding ring in a way similar to how the nuclear ring
has been proposed to be formed, (2) the merging with a low-mass
gas-rich companion galaxy, and (3) gas accretion from the galaxy halo.

In the future we plan to obtain echelle spectra of individual clumps
in the rings, which, in combination with our hydrodynamical models for
the evolution of starbursts (Silich \& Tenorio-Tagle 1998), will allow
us to understand the mechanisms that led to the formation of the
structures observed in Mrk~409. This may provide further clues to
understanding the actual impact of the recent star formation activity
on the future evolution of dwarf galaxies.

\acknowledgments

We acknowledge the referee (R$.$ Buta) for his valuable comments. We
thank the Palomar, LCO, Teide, NOT, \& CAHA observatories staff for
their help and support. This work has been financed by the Spanish PN
de Astronom\'{\i}a y Astrof\'{\i}sica grants AYA2000-1790 and
AYA2003-1676, the DFG grants FR325/50-1 and FR325/50-2, the CONACYT
(Mexico) grant 36132-E, the EC Marie-Curie grant HPMF-CT-2000-00774,
and the NASA GALEX mission.

\clearpage
\begin{figure}
\begin{center}
\epsscale{1.00}
\plotone{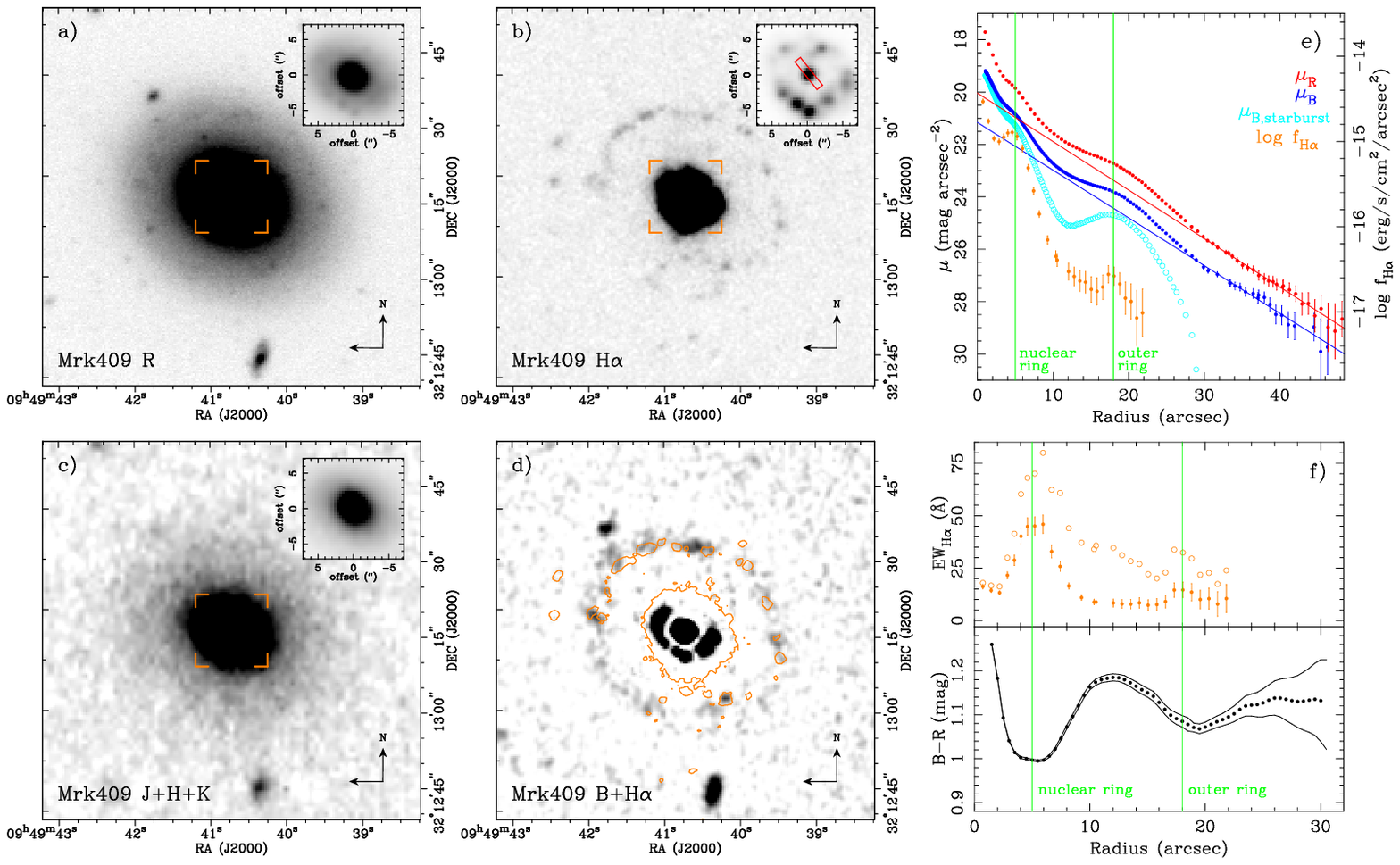}
\caption{{\bf a)} $R$-band image of Mrk~409. An expanded view of the
nuclear region is shown on the upper-right corner of each panel. {\bf
b)} Continuum-subtracted H$\alpha$ image. The image has been smoothed
using a 2$\times$2 pixels boxcar filter. The position of the slit used
for our MIKE observations is shown in the upper-right box. {\bf c)}
$J$+$H$+$K$ image. {\bf d)} Unsharp-masked $B$-band image and
H$\alpha$ contours. {\bf e)} $BR$ and H$\alpha$ surface brightness
profiles. Open circles represent the starburst $B$-band surface
brightness subtracted from the contribution of the LSB component. {\bf
f)} EW(H$\alpha$) (upper panel) and $B-R$ color (lower panel)
profiles. Filled and open circles in the upper panel represent the
starburst EW(H$\alpha$) before and after the correction from the LSB
component contamination, respectively.
\label{figure1}}
\end{center}
\end{figure}
\begin{figure}
\begin{center}
\epsscale{1.00} \plotone{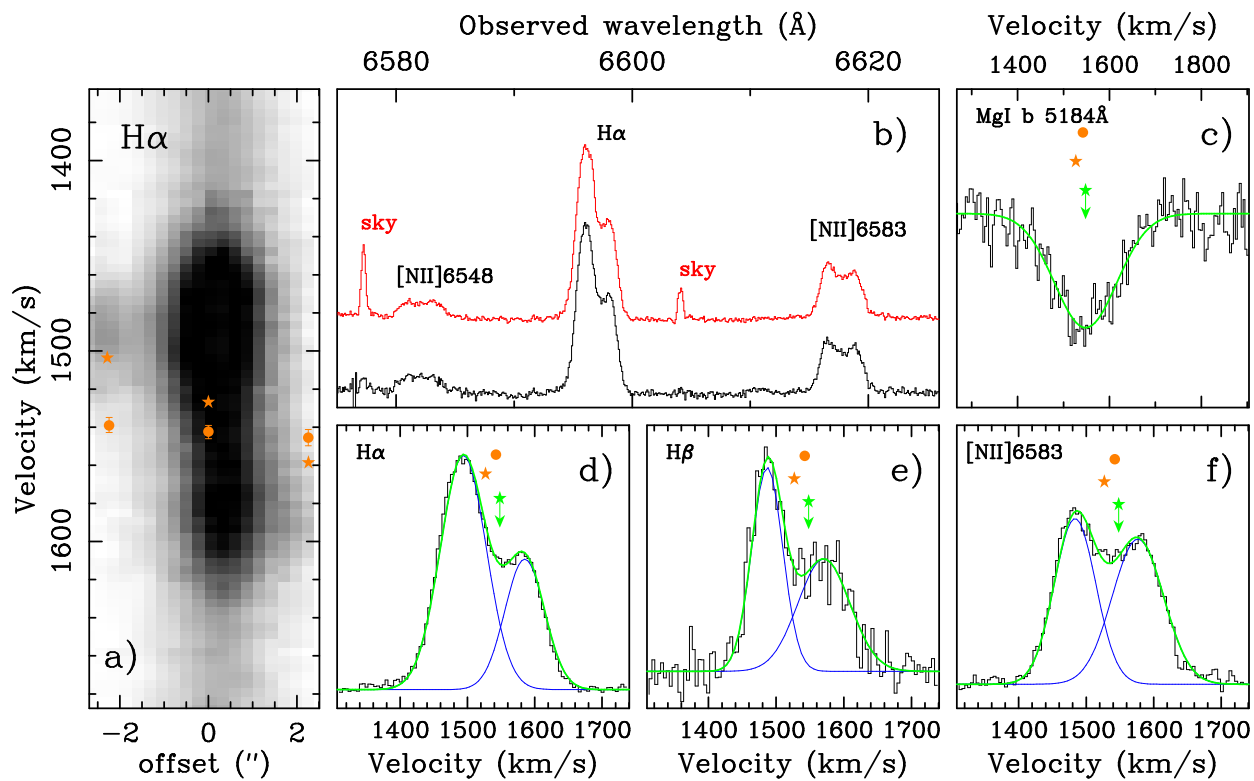}
\caption{{\bf a)} Echelle spectrum of Mrk~409 in the H$\alpha$ region
obtained with MIKE at Magellan-I. The positions and velocities for the
gas (filled circles) and stars (star symbols) published by KF01 are
shown. {\bf b)} Extracted spectrum in the region of [NII]6548\,\AA,
H$\alpha$, and [NII]6583\,\AA\ with (red line) and without the sky
(black line). {\bf c)} Extracted spectrum in the region of the MgI b
5184\,\AA\ line. The green star marks the best-fit stellar
velocity. {\bf d)} Blow-up in the H$\alpha$-line region. The
best-fitting solution using the sum of two gaussians is shown. {\bf
e)} H$\beta$. {\bf f)} [NII]6583\,\AA.
\label{figure2}}
\end{center} 
\end{figure} 

\end{document}